# Selective Parallel Loading of Large-Scale Compressed Graphs with ParaGrapher

Mohsen Koohi Esfahani
0000-0002-7465-8003

Marco D'Antonio
0009-0002-7399-7705

Syed Ibtisam Tauhidi
0000-0003-2265-2838

Thai Son Mai
0000-0002-9379-8335

Hans Vandierendonck
0000-0001-5868-9259

**https://blogs.qub.ac.uk/DIPSA**

## ABSTRACT

Comprehensive evaluation is one of the basis of experimental science. In High-Performance Graph Processing, a thorough evaluation of contributions becomes more achievable by supporting common input formats over different frameworks. However, each framework creates its specific format, which may not support reading large-scale real-world graph datasets. This shows a demand for high-performance libraries capable of loading graphs to (i) accelerate designing new graph algorithms, (ii) to evaluate the contributions on a wide range of graph algorithms, and (iii) to facilitate easy and fast comparison over different graph frameworks.

To that end, we present **ParaGrapher**, a high-performance API and library for loading large-scale and compressed graphs. ParaGrapher supports different types of requests for accessing graphs in shared- and distributed-memory and out-of-core graph processing. We explain the design of ParaGrapher and present a performance model of graph decompression, which is used for evaluation of ParaGrapher over three storage types.

Our evaluation shows that by decompressing compressed graphs in WebGraph format, ParaGrapher delivers up to 3.2 times speedup in loading and up to 5.2 times speedup in end-to-end execution (i.e., through interleaved loading and execution) in comparison to the binary and textual formats.

ParaGrapher is available online on https://blogs.qub.ac.uk/DIPSA/ParaGrapher/.

## 1 INTRODUCTION

The literature on High-Performance Graph Processing includes several graph frameworks that optimize graph algorithms for different optimization metrics. The frameworks are implemented using different parallelization libraries/techniques in different programming languages. Since frameworks have designed incompatible and different formats for the input datasets, **a comprehensive evaluation across different frameworks and for a wide range of graph datasets becomes time-consuming, inaccurate, and sometimes impossible**, especially for the emerging trillion-scale datasets such as Software Heritage version-control-history graphs[1] [4] and MS-BioGraphs sequence similarity graphs[2] [24, 26] that are published in WebGraph compressed format [5].

For those high-performance graph frameworks that are mostly implemented in C/C++, the Java implementation of WebGraph may imply that compressed graphs should be decompressed into a framework-designed format, stored in the storage, and then be accessed by the framework. **This process is complicated, error-prone, and invalidates the main benefits provided by graph compression** in WebGraph, i.e., reducing the required space on storage and the load time.

Moreover, **designing a new graph algorithm/framework requires investing time for implementing solutions for loading graphs** (as input datasets to the algorithms), while the main target of the graph frameworks is, usually, to optimize execution of graph algorithms. Due to this focus, we observe that graph processing frameworks often have highly optimized processing steps and limited optimizations for graph loading. Nonetheless, it has been argued that **evaluating the end-to-end execution time is a more appropriate metric than evaluating only the processing step** [1, 38, 40]. As such, there is also value in separately studying and optimizing the process of loading graphs from designing high-performance graph algorithms.

To this end, we present **ParaGrapher**, an API and its implementation as a library for loading graphs. ParaGrapher supports (i) **synchronous (blocking) and asynchronous (non-blocking)** loading of (ii) unweighted, vertex- and edge-weighted graphs (iii) in different formats, including WebGraph. ParaGrapher is designed to support several use cases, including **loading a graph in full, or loading specific parts of the graph**, down to a single vertex's neighbor list. This versatility helps ParaGrapher to be used in a variety of architectures such as **shared-memory, distributed-memory, and disk-based processing**.

By delegating the choice of graph formats to graph dataset publishers, ParaGrapher helps graph processing frameworks to concentrate on processing data in a more efficient way, which **accelerates and simplifies the evaluation and refinement of new (and previous) falsifiable contributions on a wider range of datasets**.

[1]https://docs.softwareheritage.org/devel/swh-dataset/graph/dataset.html
[2]https://blogs.qub.ac.uk/DIPSA/MS-BioGraphs

| Format | bits/edge | Largest Graph Size (\|E\|) |
|---|---|---|
| Matrix Market | 82.9 | 8 Billion |
| Adjacency Graph | 84.5 | - |
| Binary CSX | 32.8 | - |
| WebGraph | 13.2 | 2.5 Trillion |

**Table 1: Graph formats**

To build a theoretical basis for the evaluation of ParaGrapher (and generally for evaluation of loading compressed graphs), we construct a **performance model of loading compressed graphs** to elucidate the relative impacts of (i) storage bandwidth, (ii) compression rate, and (iii) decompression speed on graph loading time. This model is used to specify when a greater level of compression accelerates graph loading and graph loading is bounded by the decompression speed (i.e., further increase in compression ratio does not accelerate the graph loading process).

We present the design of ParaGrapher's API and the implementation of this API in ParaGrapher's library. To access compressed graph datasets, ParaGrapher provides a C/C++ front-end connected to a parallel Java back-end that decompresses the whole graph or requested subgraph using the WebGraph framework.

We empirically **evaluate the performance of ParaGrapher on three storage types (HDD, SSD, and NAS)** in comparison to graph formats (textual and binary) in GAPBS[2], a state-of-the-art graph framework for loading graphs and end-to-end execution of a graph analytic algorithm. The evaluation shows that ParaGrapher provides up to 3.2 times speedup in loading graphs and up to 5.2 times speedup in end-to-end execution.

The contributions of this paper are:

- Introducing the ParaGrapher API and library for loading large-scale graphs,
- Modeling decompression efficiency based on the storage characteristics, and
- Evaluating ParaGrapher on three types of storage and in comparison to a state-of-the-art graph framework.

The paper is structured as follows. Section 2 reviews the background of graph storage formats. Section 3 presents a model for decompression efficiency. Section 4 explains the design of ParaGrapher API and its implementation for WebGraph format. Section 5 empirically evaluates ParaGrapher. Section 6 presents an analysis and future directions. Section 7 reviews further related works and Section 8 concludes the discussion.

## 2 BACKGROUND

**Terminology**. A graph $G = (V, E)$ has a set of vertices $V$, and a set of directed edges $E$. The adjacency matrix is a binary matrix representing the graph: the element at row $i$ and column $j$ is 1 if $E$ contains an edge from vertex $i$ to $j$, and 0 otherwise. For an edge-weighted graph, the elements of the adjacency matrix show the weights on the corresponding edges between vertices.

**Graph Loading**. Graph processing starts by preparing access to the input dataset, which is either (i) read from the secondary storage, (ii) built by a synthetic graph generator [7], (iii) extracted from table(s) in relational database(s), or (iv) computed from other graph formats such as sum graphs [15] or context-free grammars [42]. In this paper, we focus on the first method, loading from storage, which is widely used in High-Performance Computing.

**Textual and Binary Formats**. Storage-based graph loading consists of accessing storage for reading data represented in a particular format. The *Textual Coordinated* (Txt. COO) or *Matrix Market* [3] is used in graph collections such as Network Repository[3] [48], Konect[4] [35], and SuiteSparse[5] [11]. The input file contains a number of lines, where each line presents the ID of the endpoint vertices of an edge. The *Textual Metis* format [20] stores the neighborhood of each vertex in the file. The *Textual Adjacency* (Txt. CSX) format [55] represents the Compressed Sparse Row (CSR) or Column (CSC) [51] format of the graph as a textual file. The Compressed Sparse format consists of two arrays: (1) an *offsets* array containing $|V| + 1$ elements, and (2) an *edges* array of $|E|$ elements. The offsets array is indexed by a vertex ID and specifies the index of the first edge of that vertex in the edges array. The edges array specifies the ID of the source endpoint of the edges for the CSC and the destination endpoint for the CSR.

While a textual format is easy for humans to read, it wastes space and requires text-to-binary conversion during reading, which increases loading time. To prevent this, *binary* presentation of the above formats may be used.

**Compressed Formats**. To better use disk space, the stored graph may be compressed. Graph compression is also vital for the efficient transferring of graph datasets over a network. WebGraph[6] [5] is the most long-lasting (over 25 years) and accurate open-source graph compression and processing framework, implemented mainly in Java. The framework had been mainly targeted for Web graphs where vertices represent URLs and edges show the hyperlinks, however, it is now used for compressing and processing different types of real-world graphs [4, 26].

WebGraph compression works on top of two features: (i) Locality: most links refer to pages within the same host, and (ii) Similarity: pages that are close to each other in lexicographic order usually have common successors. Graphs are compressed in WebGraph format using delta-encoding, reference compression, differential compression, and interval representation techniques. In this way, WebGraph may facilitate up to 35× compression.

Table 1 shows (i) a comparison of #bits/edge for different formats of graphs used in this paper (Table 3) and (ii) the size of largest (known to us) public graph in each format.

**Parallel Loading**. All formats mentioned above have the potential to be loaded in parallel. The file contents of a Textual COO graph can be divided into a number of chunks that are assigned to threads to be processed. Each thread first specifies the number of edges in its chunk. The prefix sum of the number of edges in chunks will be used in the second pass to specify the index for writing the edge. Similarly, the textual CSX and Metis formats can be read in parallel. Binary formats can be read more easily by dividing the file's total size between threads, and each thread reads its chunk in parallel with the others.

**Parallel Decompression**. WebGraph provides random access to different parts of a compressed graph and parallel threads can start concurrently accessing different parts of the graph and decompressing it. We explain the details of this procedure in Section 4.

[3]https://networkrepository.com/
[4]http://konect.cc/
[5]https://sparse.tamu.edu/
[6]https://webgraph.di.unimi.it/

## 3 MODELING COMPRESSION EFFICIENCY

In this section, we present a mathematical model to shed light on the opportunities and limitations of loading compressed graphs. We assume that the storage system has a fixed average read bandwidth of $\sigma$, measured in GigaBytes per second (GB/s). So, the uncompressed graph data (stored in a binary representation that matches one-to-one with the in-memory graph layout) is transferred from storage to memory at a rate of $\sigma$ GB/s.

The compressed graph is transferred from storage to memory at the same rate. However, the transferred graph should be decompressed, which requires computation. Working in a multi-processing environment, different CPU cores are involved in different tasks, and an extensive overlap between computation and data movement happens. The overlapping of computation and data movement results in two scenarios:

(i) **Decompression takes relatively little time, which is fully overlapped with data movement**. In this scenario, the data movement (i.e., storage bandwidth) is the main factor in specifying the throughput bound. Assuming a compression ratio of $r > 1$ (i.e., $r$ Bytes have been compressed into 1 Byte), an upper bound on the rate at which a compressed graph can be loaded is $\sigma r$.

(ii) **Decompression takes relatively longer than data movement**. In this case, graph loading is bounded by the decompression bandwidth $d$. Both $r$ and $d$ depend on the compression algorithm, but $d$ may vary with compression rate $r$, and the speed of the CPUs (i.e., frequency and number of cores). In fact, by increasing $r$, $d$ is reduced, but the largest value of $d$ remains the upper bound. We ignore dependencies on CPU speed and compression algorithm. The second upper bound for graph loading is, thus, identified by $d$.

Combining the above bounds, **the graph loading b has a lower bound $\sigma \le \mathbf{b}$ and an upper bound $\mathbf{b} \le \min(\sigma \mathbf{r}, \mathbf{d})$**.

Figure 1 shows the load bandwidth for two storage types: (i) HDD with 160 MB/s average read bandwidth and (ii) SSD with 3.6 GB/s average read bandwidth. The quantities 160 MB/s and 3.6 GB/s are bandwidth measured in this work, which is explained in Section 5.1.

Figure 1 shows that when **using a storage with a high throughput (such as SSD), it is necessary to have faster decompression methods** as it is the computation limit ($d$) that specifies the efficiency of the decompression process. On the other hand, **for a low-bandwidth storage (such as a single HDD or a remote storage bounded by network limits and shared between users), compression efficiency is mainly limited by the graph compression ratio**.

We use this model in Sections 5 to evaluate the throughput of ParaGrapher on different storage types and in Section 6 to analyze the decompression process.

## 4 PARAGRAPHER

Section 4.1 reviews the requirement of a high-performance graph loading library to be used effectively in different applications. Section 4.2 explains the high-level design of the ParaGrapher API, which is detailed in Section 4.3. Section 4.4 explains the implementation of the ParaGrapher for graphs in WebGraph format.

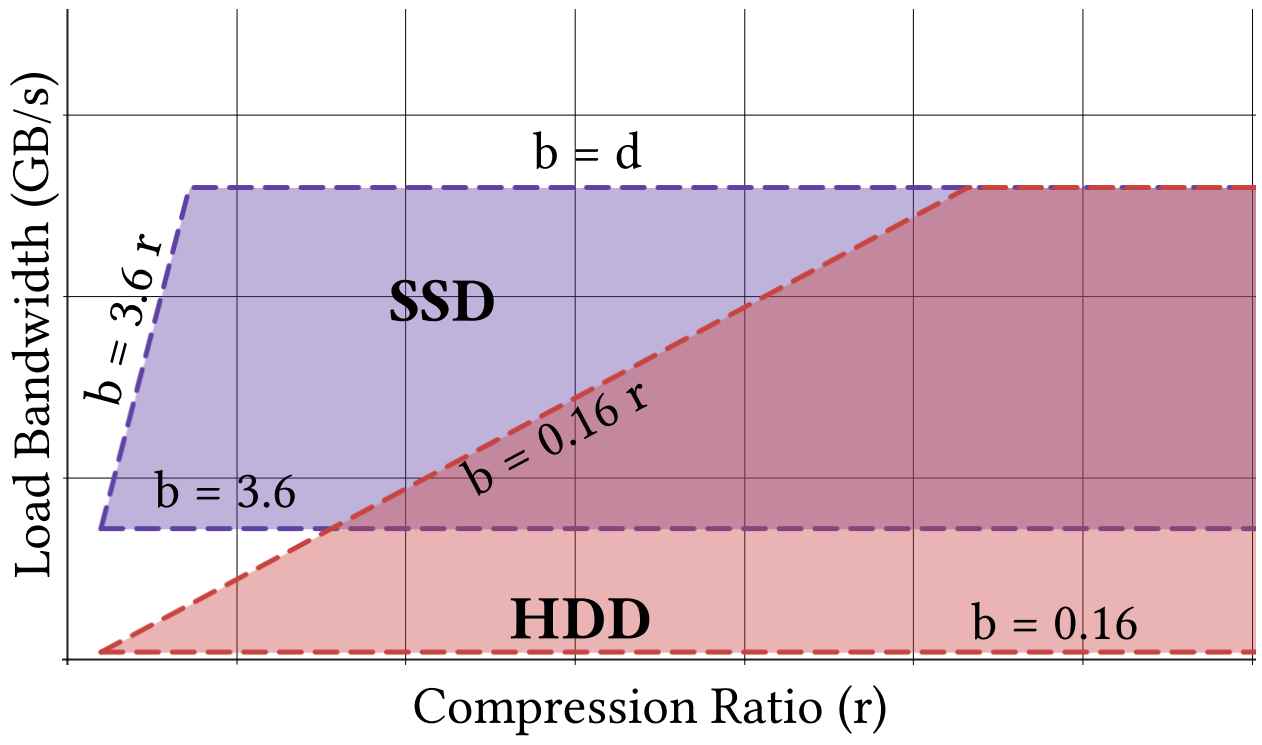

**Figure 1: Load bandwidth model for HDD (with $\sigma = 160MB/s$ bandwidth) and SSD ($\sigma = 3.6GB/s$ bandwidth)** - The X-axis shows the Compression Ratio ($r$), and the Y-axis shows the Load Bandwidth ($b$). The decompression bandwidth (computation limit) is shown by $d$. $b$ is modeled as $\sigma \le b \le min(\sigma r, d)$.

### 4.1 Use Cases and Design Requirements

Static graphs are accessed by different shared-memory, storage-based, and distributed-memory frameworks in 4 ways:

- **A.** Each edge may be accessed more than one time in algorithms such as Breadth-First Search and Label Propagation Connected Components. This is a prevalent access pattern in **shared-memory** graph frameworks.
- **B.** Each edge is accessed once and processed independently from the other edges in algorithms such as Jayanti-Tarjan Connected Components (JT-CC) [19] and Low-Diameter Decomposition (LDD) [45].
- **C.** A subgraph containing a number of consecutive blocks of edges is accessed by each machine in a **distributed-memory** graph processing framework [43, 44, 53].
- **D.** All edges are required, but memory cannot store all edges. Therefore, periodically, blocks of consecutive edges are loaded from storage and processed. This is used in **storage-based (out-of-core)** graph frameworks [37, 50, 59].

In addition to support the above access types, it is necessary for a graph loading API to support parallelism. Computing clusters are equipped with high-bandwidth parallel storage/file systems and support concurrent storage accesses. Moreover, it is necessary for the highly compressed datasets to be decompressed in parallel as the decompression may require a longer time than loading the compressed graph. As such, **the underlying compression format should support parallel and random access**.

Memory management should also be planned in a graph loading API as graph algorithms are memory-intensive and simple assumptions such as allocating memory for all requested edges may not be applicable (use case D) or even required (use case B). So, **it is necessary for the API to support solutions that minimize the impacts that may inherently be implied into the memory management layer** to allow the user application/framework to have a full authorization around arranging memory.

Minimizing the impacts made on computational resources by calling a function from a graph loading library is not limited to memory allocation for edges. The **impacts on RAM-based OS caching of storage contents and finalizing/killing threads (which have been started by the library) should also be considered**. By loading (and decompressing) data from the storage, the OS may cache storage contents in RAM and if this cache is not dropped, the next memory allocations in the main program/framework may experience some delays. On the other hand, the library should ensure the created threads for load parallelization are killed/finished and do not consume CPU cycles after completion of the load process. [7].

## 4.2 ParaGrapher API Design

To support all four use cases enumerated in Section 4.1, ParaGrapher uses the **finest granularity** as its base, and the user asks for a **consecutive block of edges**. If the whole graph is required, the user passes the whole range as input to ParaGrapher.

The ParaGrapher API supports loading graphs in **blocking and non-blocking** modes. The implementing library may block the caller thread and return only after completion of loading the graph or requested subgraph. Figure 2 shows an example of a synchronous (blocking) call.

Alternatively, the user may need to **overlap loading graph with other tasks** such as computation and communicating with other machines (particularly for use cases B, C, and D). In this case, a call for loading a subgraph is performed asynchronously, i.e., in a non-blocking mode and user specifies a *callback* that is invoked by the library when all or part of the requested edges have been read. Figure 3 shows an example of an asynchronous (non-blocking) call. The call is returned immediately to the user after creating a thread to load the requested edges. Loading is divided into a number of edge blocks that are perfromed in parallel, and the callback is invoked on a new thread after the completion of loading each block.

In the ParaGrapher API, the user may allocate memory for the block of edges and pass it to the graph loading function. The API also supports **storing data in reusable buffers allocated and managed by the library**. The buffers are then passed to the user, who is responsible for transferring data to its own data structures. In this way, ParaGrapher minimizes applying restrictions to the memory management layer of the graph framework.

## 4.3 ParaGrapher API in Details

In this section, we review the ParaGrapher API. A more detailed list of functions is presented in Appendix A and online on ParaGrapher repository[8]. Function names start with `paragrapher_`, which is skipped for the sake of brevity.

To initialize the library, `init()` is called and for opening a graph `open_graph()` is used, which receives the graph's path and type. To release an opened graph, `release_graph()` is used. The `get_set_options()` function is called to query (get) the properties of the graph or to set some options of the library. This function can be used to receive the number of vertices, edges, if the library reads the graph synchronously or asynchronously, getting/setting the buffer size of threads and the number of buffers, and to query if loading a graph is completed or how many edges have been read.

[7]E.g., if the graph loading library uses OpenMP and an "active" value for `OMP_WAIT_POLICY`, and if the main graph processing framework is run on the same process but using another parallelization library, the OMP threads may continue consuming CPU cycles.

[8]https://github.com/DIPSA-QUB/ParaGrapher/wiki/API-Documentation

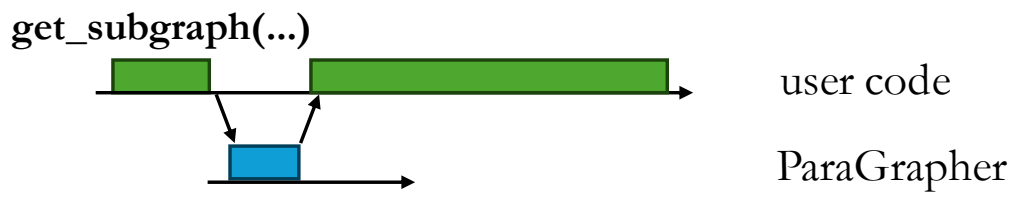

**Figure 2: A synchronous (blocking) call.** The library parallelizes loading/decompression while user is waiting for receiving the requested graph/subgraph **all at once**.

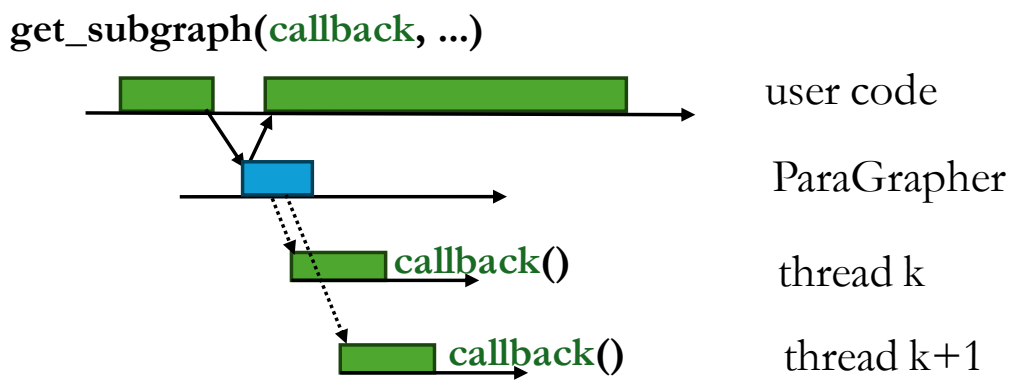

**Figure 3: An asynchronous (non-blocking) call.** The execution is immediately returned to the user and the library parallelizes the loading/decompression. Upon completion of loading/decompressing each block of edges, the user is notified through the `callback()`. In this way, the user **gradually** receives the requested graph/subgraph and may start processing them as soon as the first block of edges has been returned.

| Type | Vertex ID Size (Bytes) | Vertex Weight Size (Bytes) | Edge Weight Size (Bytes) |
|---|---|---|---|
| `CSX_WG_400_AP` | 4 | 0 | 0 |
| `CSX_WG_800_AP` | 8 | 0 | 0 |
| `CSX_WG_404_AP` | 4 | 0 | 4 |

**Table 2: WebGraph types in ParaGrapher**

To get offsets of vertices of a graph in CSX format, the function `csx_get_offsets()` is used, and to get vertex weights function `csx_get_vertex_weights()` is used.

To get a block of edges (and their weights if the graph is edge-weighted) in CSX format, function `csx_get_subgraph()` is called. Based on the type of access (synchronous/asynchronous) implemented by the library, this function can be called in both synchronous (i.e., blocking) and asynchronous (i.e., non-blocking) modes. If the library passes its internal buffers to the user, the user is required to call `csx_release_read_buffers()` after it has finished accessing the buffer. Similarly, for the graphs in COO format, the function `coo_get_edges()` is used.

## 4.4 WebGraphs in ParaGrapher Library

ParaGrapher supports compressed graphs in WebGraph format and categorizes them into three types shown in Table 2.

The type names end with two characters: the first one is either `A` or `S` and indicates if the graph is loaded asynchronously or synchronously. The second character is `P` or `S` to show if the library loads in parallel or in serial. Note that the current edge-weighted graphs have less than $2^{31}$ vertices, however, ParaGrapher can easily be extended to support other weights such as `WG_80X_AP`, X={4,8}.

For loading a WebGraph, ParaGrapher implements a consumer-producer pattern where the consumer side is implemented in the C language and the producer side is implemented in Java (as the main language of the WebGraph framework). ParaGrapher uses shared memory to communicate between the sides. The C side program allocates the shared memory for (i) the buffers that contain the read edges and (ii) for the metadata of each buffer that contains the status of the buffer and the start and end vertex and edges. A buffer can be in one of the 5 statuses:

- `C_IDLE`: the buffer is ready to be allocated for reading an edge block.
- `C_REQUESTED`: the metadata of the buffer has been set, and the Java program can start reading the buffer.
- `J_READING`: the Java program has created a new thread for reading the buffer.
- `J_READ_COMPLETED`: reading has been completed by the Java program.
- `C_USER_ACCESS`: the buffer is accessed by the user and the library cannot use it until the user releases it.

When a buffer is in `C_IDLE` status and a block of edges should be read, the C program sets the metadata of the related buffer and changes its status to `C_REQUESTED`. The Java program observes there is a new request, creates a thread for reading the requested block of edges, and sets the status of the buffer to `J_READING`. When reading is completed, the Java program sets the status to `J_READ_COMPLETED` to inform the C program that the block has been read and the user can access the buffer. By observing this status, the C program sets the status of the buffer to `C_USER_ACCESS` and creates a new thread to run the user-defined callback function. By the end of the callback function, the user releases the buffer, and its status is set to `C_IDLE` so that the C program can start requesting a new block of edges.

Both sides access the status field in buffer metadata, but in each step, based on the current value, it is modified by only one side, and the other side only observes it. So, the observer will see either the old or new values. By observing the new value, the observer starts the required operation on its side. Since the observer starts (and does not stop) an operation after observing a new value in the status field, delays in propagation of the changed status from CPU register to cache and then to memory until it is observed by the observer thread (running on another CPU core) does not affect the correctness of the synchronization. However, the modifier thread should ensure that the state modification happens as the last modification to the buffer and its metadata.

In Java side this ordering is guaranteed by the program-order happens-before semantics of the Java Memory Model for instructions executed by a single thread. In C side the ordering is provided using explicit memory barrier[9].

As we explain in Section 5.5, the ParaGrapher's WebGraph library creates up to $2 \times$ `#cores` concurrent threads for parallelizing the loading process. The default value of buffer size is 64 Million edges. The user may change these values. The library tracks the requests and sends new requests when the previous buffers are free.

[9] `__sync_synchronize()` in GCC

| Abbr. | Name | \|V\| | \|E\| | Size on Storage | | | |
|---|---|---|---|---|---|---|---|
| | | | | Txt. COO | Txt. CSX | Bin. CSX | WebGraph |
| RD | US Roads[12] [48] | 23 M | 58 M | 940 MB | 671 MB | 403 MB | 122 MB |
| TW | Twitter 2010[13] [36] | 42 M | 2.4 B | 40 GB | 21 GB | 9.3 GB | 4.1 GB |
| G5 | Graph500 RMAT[14] | 540 M | 16 B | 310 GB | 168 GB | 68 GB | 51 GB |
| SH | SWH Gitlab[15] [4] | 1 B | 55 B | 1.0 TB | 512 GB | 217 GB | 26 GB |
| CW | ClueWeb 2012[16] | 1 B | 74 B | 1.4 TB | 701 GB | 286 GB | 17 GB |
| MS | MS50[17] [26, 34] | 585 M | 124 B | 2.3 TB | 1.3 TB | 470 GB | 325 GB |

**Table 3: Datasets** - M: Million, B: Billion, MB: MegaBytes, GB: GigaBytes, TB: TeraBytes

In this way, a queuing system is not required for communication between the C and Java sides.

To ensure **ParaGrapher returns the computational resources as they were before calling** (Section 4.1), (i) ParaGrapher finishes all threads it creates in its Java-side and tracks the status of threads it creates in C-side for the `callback` specified by the user, (ii) ParaGrapher minimizes its memory impacts through sharing its buffers with user and delegating transfer to the user, and (iii) the OS cache of storage contents is evicted through OS interfaces[10] or through calling the `flushcache` program[11] in the ParaGrapher library with the same functionality.

## 5 EVALUATION

**Experimental Setup**. We use a machine with an Intel W-2295 3.00GHz CPU, 18 cores, 36 hyper-threads, 24MB L3 cache, 256 GB DDR4 2933Mhz memory, running Debian 12 Linux 6.1. The machine has a 4TB Samsung MZQL23T8HCLS-00A07 PCIe4 NVMe v1.4 SSD and a 6TB Hitachi HUS726060AL 7200RPM SATA v3.1 HDD. The storage is used without any RAID or caching mechanism.

We also measure the performance of ParaGrapher on a TS-853DU-RP NAS containing 4×ST16000NM001G 16TB Seagate HDDs connected through a switch to a cluster. A machine with 2 Intel Xeon Gold 6438Y+, in total, 64 cores, 128 hyper-threads, 120 MB L3 cache, and 256 GB memory used for NAS experiemnts.

We used `gcc` 12.2, OpenMP [10] for C parallelization, `OpenJDK` 17.0.10, and `WebGraph` 3.6.10. We use GAPBS[18] [2] as baseline comparison. GAPBS is a state-of-the-art graph framework that supports different graph formats.

Table 3 shows the graph datasets we use for evaluation. `G5` is a RMAT synthetic graph ($log_2|V| = 26, log_2|E| = 33$) and the other ones are real-world graphs. We compressed the binary versions of `G5` and `RD` to WebGraph format using the WebGraph framework. Compressing these two graphs required 6.4 and 2,480.1 seconds, respectively. The other datasets have only been published in WebGraph format, and we symmetrized the asymmetric ones. For comparison with GAPBS, we created CSX Binary and COO edge list presentation of the graphs. The graphs are encoded with 4 Bytes ID per vertex as $|V| < 2^{32}$. In the CSX format, the `offsets` array requires 8 Bytes per entry as $|E| > 2^{32}$ for some datasets.

[10] E.g., `/proc/sys/vm/drop_cache` in Linux, https://www.kernel.org/doc/Documentation/sysctl/vm.txt.
[11] https://github.com/DIPSA-QUB/ParaGrapher/blob/main/test/flushcache.c
[12] https://networkrepository.com/road-road-usa.php
[13] https://law.di.unimi.it/webdata/twitter-2010/
[14] https://github.com/graph500/graph500/tree/newreference/generator, commit 4ff7573
[15] https://docs.softwareheritage.org/devel/swh-dataset/graph/dataset.html#gitlab-all
[16] https://law.di.unimi.it/webdata/clueweb12/
[17] https://doi.org/10.21227/gmd9-1534
[18] https://github.com/sbeamer/gapbs/, commit 33f73f4

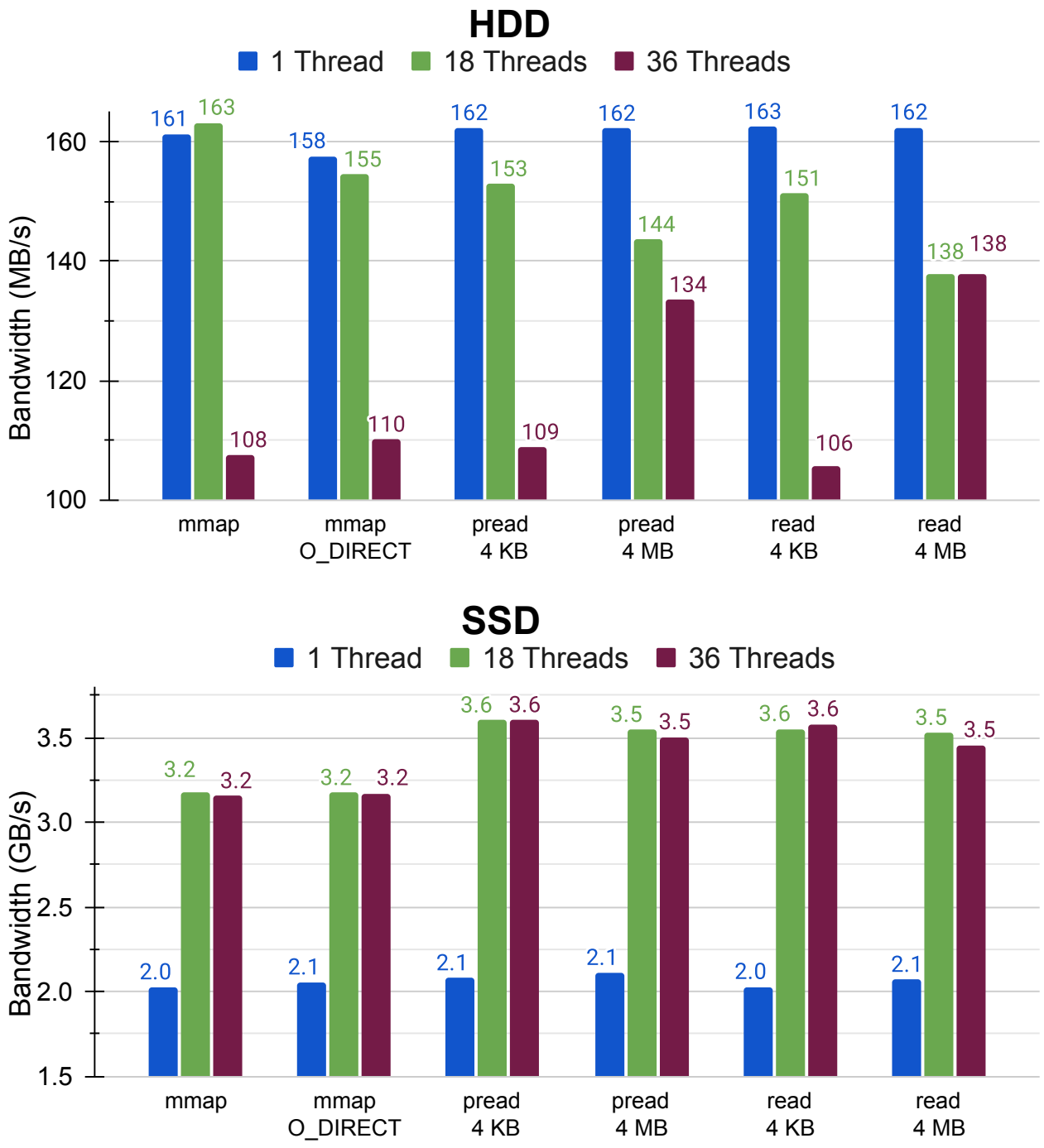

**Figure 4: HDD and SSD read bandwidth benchmarking**

## 5.1 SSD and HDD Read Bandwidth

To have a correct evaluation of the efficiency of the compression, it is required to have the baseline bandwidth of different storage types. To that end, in this section, we present an evaluation[19] of the SSD and HDD read bandwidth for (i) different sizes of blocks (4KB and 4MB), (ii) for 1, 18, and 36 threads, and (iii) for different read methods: `mmap`[20], `pread`[21], and `read`[22]. In this experiment, a 12GB file was used for each evaluation, and the OS cache of storage contents was dropped before the evaluation. The content of the file has been divided between the threads based on the block size granularity (i.e., 4KB/4MB). In the case of `mmap`, we report accessing file opened with/out `O_DIRECT`.

Figure 4 shows the results of this evaluation. It shows that **the bandwidth of HDD can be saturated using a single thread; however, for SSD, a greater number of threads are required**. Moreover, in HDD, by increasing the number of threads, we may see degradation in bandwidth. Figure 4 also shows that `mmap` reduces the bandwidth of SSD and using `O_DIRECT` does not have much impact on `mmap`. In total, we can practically expect **a 160 MB/s bandwidth for HDD and a 3.6 GB/s bandwidth for SSD**.

[19]https://github.com/DIPSA-QUB/ParaGrapher/blob/main/test/read_bandwidth.c
[20]https://man7.org/linux/man-pages/man2/mmap.2.html
[21]https://man7.org/linux/man-pages/man2/pread.2.html
[22]https://man7.org/linux/man-pages/man2/read.2.html

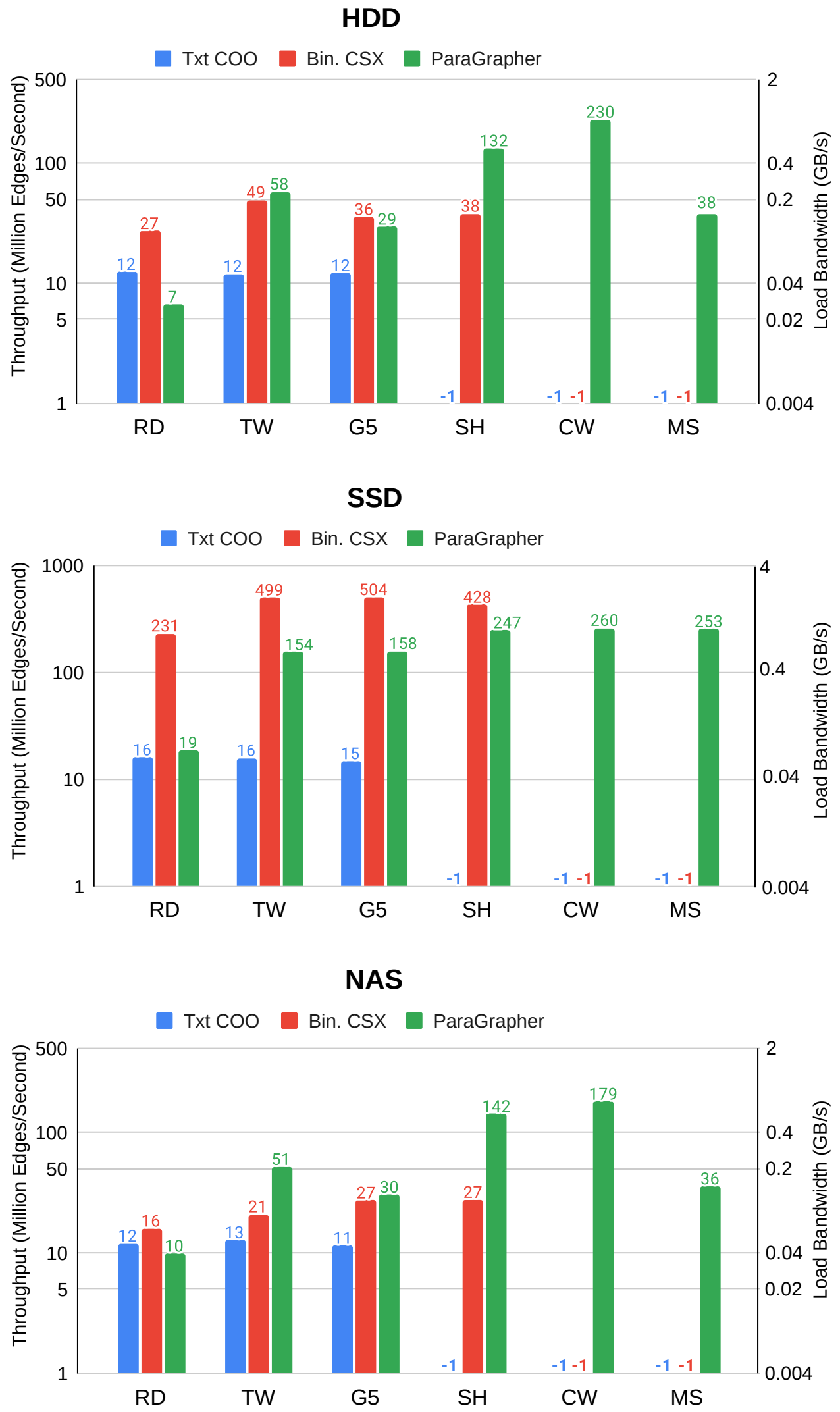

**Figure 5: Throughput and Load Bandwidth of ParaGrapher vs. textual and binary formats.** The left Y-axis shows throughput in Million Edges per Second. The right Y-axis shows Load Bandwidth in MB/s and GB/s. Numbers on bars are Throughput values (the left Y-axis). "-1" indicates an "Out of Memory" error.

## 5.2 Graph Loading

In this section, we evaluate the performance of ParaGrapher in comparison to Binary CSX and Textual COO formats used in GAPBS for loading graphs. We run the experiments for HDD, SSD, and NAS, which are shown in Figure 5.

Section 5.1 shows that the HDD storage has a bandwidth of 160 MB/s, and with 4 Bytes/edge we can expect a throughput of around 40 Million Edges/second (ME/s) for an uncompressed binary graph in CSX format. The HDD plot in Figure 5 shows that **ParaGrapher delivers a throughput of up to around 129 ME/s, i.e., 3.2 times the base storage throughput**, which indicates an efficient usage of compression. Also, we see that **ParaGrapher is able to load all graphs as it partially loads the graph, even if the graph cannot be stored completely in the memory**.

Similarly, we can expect a throughput of around 900 ME/s for the SSD storage, however, Figure 5 shows that ParaGrapher provides up to 260 ME/s throughput. As discussed in Section 3, load bandwidth is limited by the compression ratio, storage bandwidth, and the decompression process bandwidth. For a high-bandwidth storage, the bandwidth of the decompression specifies the limit. This shows that **for high-bandwidth storage (such as SSD), further improvements on decompression are necessary to benefit from decompression**. For SSD, we also see that the Binary CSX format provides a throughput of around 504 ME/s, which is compatible with the results in Figure 4 (Single threaded read on SSD provides around 2–2.1 GB/s bandwidth).

Figure 5 shows that ParaGrapher achieves a throughput of up to 179 ME/s when using NAS which is 7.3 times of the throughput of the Binary CSX format.

## 5.3 Partial Graph Processing

In Section 4.1, we enumerated four use cases that a graph loading library should support. In Section 5.2, we evaluated ParaGrapher for loading the whole graph (use case A). In this section, we evaluate ParaGrapher for use cases B, C, and D (storage-based and distributed-memory graph processing), where a partition of edges is required. To materialize this condition optimally, we use a graph algorithm that requires one pass over the edges and processes each edge independently from the others. By gradually loading edges and processing them, the experiment evaluates the efficiency of the ParaGrapher for use cases B, C, and D.

We compare ParaGrapher and GAPBS for the Weakly-Connected Components (WCC) graph algorithm. GAPBS includes Afforest [58] and we use an implementation[23] of Jayanti-Tarjan WCC (JT-CC) [19] with ParaGrapher. JT-CC works based on mapping a tree to each component of graph and requires one pass over edges where each edge is processed independently from other edges. In this way, JT-CC can be applied on the graph even if the graph cannot be completely stored in the memory. We use this experiment to evaluate how ParaGrapher provides partial access to a graph.

Figure 6 compares execution of WCC for different formats in GAPBS (Textual COO and Binary CSX) and ParaGrapher (WebGraph). It shows that for all graphs and for all storage types, **ParaGrapher is able to load the graph partially, pass the subgraph to the user's callback to be processed, and continue with the next partition without storing the whole graph in memory**. Figure 6 shows that ParaGrapher provides up to 5.2 times speedup, however, as we explained in Section 5.2, for graphs in Binary CSX format and for loading from SSD, the decompression bandwidth limits the performance of ParaGrapher.

## 5.4 Decompression Bandwidth

In Section 3, we explained that the load bandwidth depends on the compression ratio, graph decompression bandwidth, and on the storage bandwidth. To have a better illustration of the last two factors, in this section, we evaluate the performance of ParaGrapher

[23]https://github.com/DIPSA-QUB/ParaGrapher/blob/main/test/test2_jtcc_WG400.c

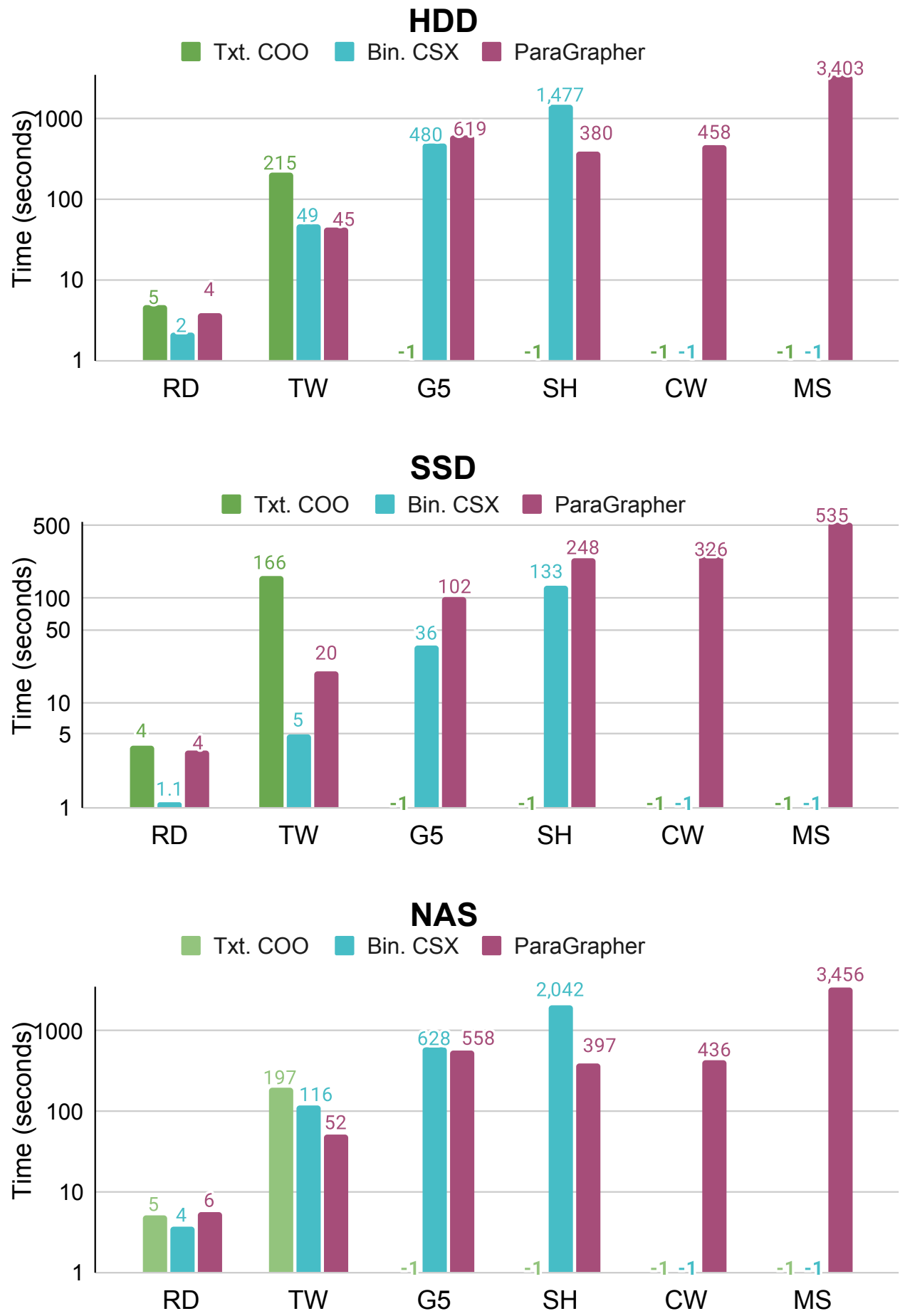

**Figure 6: Performance of Weakly-Connected Components in seconds. "-1" indicates an "Out of Memory" error.**

on different mediums. In addition to HDD and SSD that we had used in the previous sections, we use:

- **NVMM**: 1TB 3.2GHz Intel NMB1XXD128GPS non-volatile memory DIMMs on a machine with 2 Intel 5318Y CPUs, in total 48 cores, 96 threads, and 256GB memory,
- **DDR4, 3.2GHz**: 2TB 3.2GHz DDR4 memory on a machine with 2 AMD 7702, in total, 128 cores, 256 threads.

Figure 7 compares the throughput of ParaGrapher for the different storage mediums. It shows that the largest throughput for ParaGrapher is 952 Million Edges/second or 3.8 GB/s. In Section 6, we present solutions for optimizing decompression bandwidth.

## 5.5 ParaGrapher Parameters

As we explained in Section 4.4, ParaGrapher uses two parameters for parallelizing reading graphs in WebGraph format: (i) the number of buffers that specifies the number of parallel threads, and (ii) the size of each buffer that specifies the work performed by each thread. To evaluate the impacts of these parameters, we set the number of threads to 9, 18, and 36 threads and buffer sizes to 8, 16, and 128

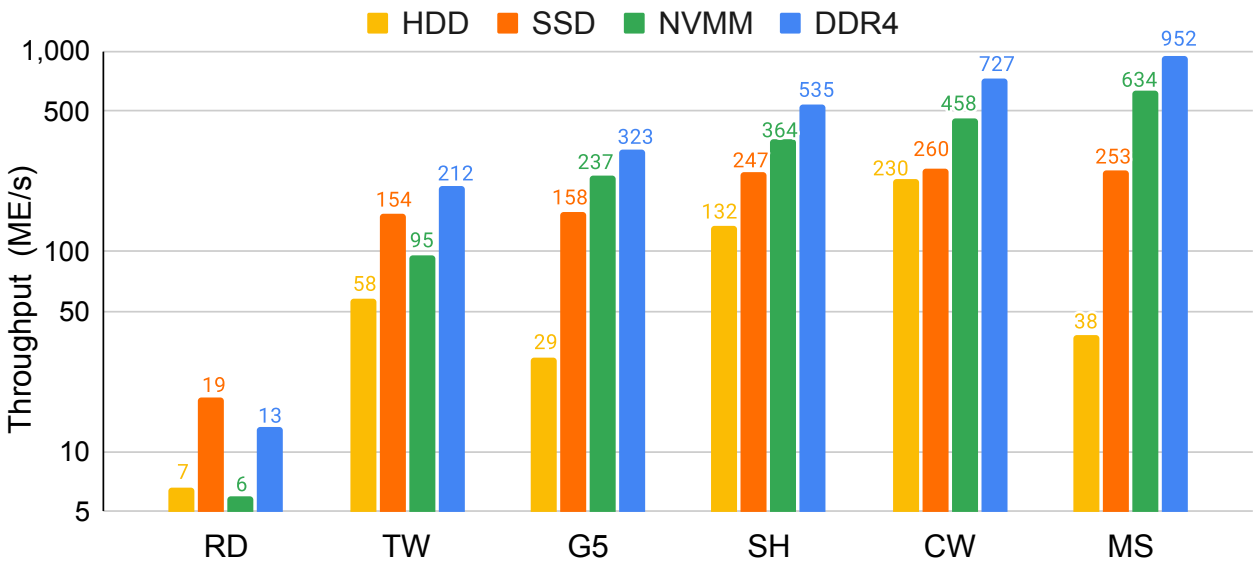

**Figure 7: ParaGrapher throughput (in Milion Edges per second) for different medium**

Million edges and consider the performance of ParaGrapher for reading graphs.

Figure 8 shows that when **using HDD, by increasing the number of threads, we may see performance degradation. However, a small number of threads reduces the performance of SSD**. Moreover, increasing the size of buffers reduces the number of total partitions that should be read. In this way, **the large size of buffers results in load imbalance between threads** and the maximum bandwidth of the load cannot be reached.

The Java-side scheduler thread periodically checks if reading buffers have been finished. As a result, if each thread has a small amount of work to do, a large fraction of time is passed waiting for the scheduler to assign another block. So, reducing the buffer size to very small values may result in performance degradation while providing more load balance.

Based on our analysis in this section, ParaGrapher uses blocks of 64 Million edges and sets the number of threads to $\#cores$ for HDD and to $2 * \#cores$ for SSD. The user can modify these parameters before starting to read the graph.

On the other hand, load performance is bounded by (i) the storage bandwidth and (ii) the decompression bandwidth (Section 3). Both factors are affected by the parallelism. Increasing the number of threads reduces the read bandwidth of HDD but is useful for SSD (Section 5.1). On the other hand, the decompression bandwidth is increased by increasing the number of threads. This shows that **the storage type, machine's computation throughput, and compression ratio specify the decompression process as either a computation-bound or a storage-bound problem**.

## 5.6 Decompression Scalability

For low-bandwidth storage, the performance of graph loading depends on the storage, but graph decompression is a computation-bound problem when using high-bandwidth storage. To measure the scalability of ParaGrapher, we use the machine with 128 cores and 2TB DDR4 3.2GHZ and store the datasets on memory. In this way, the execution is independent of the storage delays, and we can measure the scalability of decompression.

Figure 9 shows up to 3.8 times speedup using 128 cores in comparison to 16 cores. To consider the limited scalability, we need to study how a graph is loaded in WebGraph. This process consists of two steps: (i) the graph metadata is **sequentially** loaded to memory by `ImmutableGraph.loadMapped()` and (ii) graph is accessed by **parallel** threads. **A source of limited scalability in ParaGrapher is the sequential loading of graph metadata**, i.e., the first step. Our measurements shows that 12.9–60.6% of execution time is passed in the sequential step.

# 6 A POSTERIORI ANALYSIS & FUTURE DIRECTIONS

**Trading-off Between Decompression Bandwidth and Compression Ratio**. The compressed graph loading model (presented in Section 3) shows that the speed of loading compressed graphs is influenced by investing in hardware (storage systems and/or compute power) to increase $s$ and $d$. Alternatively, there is scope to evaluate and to design compression algorithms with different trade-offs for the compression rate and decompression time, i.e., finding light-weight decompression algorithms with high decompression bandwidth ($d$) that achieve high compression rate ($r$) at the same time.

**Considering Storage Types**. Section 5.1 shows that storage types such as HDD and SSD reveal different bandwidth depending on block size and number of concurrent requests. Moreover, another study[24] on LustreFS shows that O_DIRECT may improve the bandwidth by 2.15 times by removing client-side cache. This shows that using the easier solutions (such as `mmap()`) is not always the key to optimising the bandwidth and decompression process can benefit from considering the requirements of the medium for saturating its bandwidth (Section 3).

**Optimizing Storage Bandwidth**. In Section 5.5 and Section 5.6 we saw that storage bandwidth is one of the limiting factors of the decompression process. To optimize storage bandwidth, it is required to use the best read methods that consider the storage type (Section 5.1). In Figure 10, we compare the bandwidth of SSD and HDD for memory mapping and block-based reading and in C[25] and Java[26]. It shows that the Java implementation may reach 78-101% of the performance of the C implementation for reading from the storage. While WebGraph is being implemented in lower-level languages [16] to reduce overheads, language-dependent optimization may be useful for saturating the storage bandwidth. A filesystem for prefetching and caching large block sizes (e.g., 64 MB) from storage may also be helpful to prevent sending a large number of small requests to the storage[27].

**Loading From High-Bandwidth Storage Instead of Processing**. Using high-bandwidth storage (Section 5.1), we may reduce the preprocessing time by loading data of size $O(|V|)$ from storage instead of paying the processing cost of $O(|E|)$.

As an example, some graph algorithms require the transposed version of the input graph. By loading the `offsets` array of the transposed graph from the storage (that requires a load of $O(|V|)$ size), only one pass over edges is required for creating the transposed graph, and the $O(|E|)$ computation for creating `offsets` is prevented. Some recent datasets include the `offsets` array [25, 26].

Also, WebGraph requires random access to the underlying graph to extract the degree of vertices. If the `offsets` array of the CSX

[24] https://blogs.qub.ac.uk/DIPSA/HDD-vs-SSD-vs-LustreFS-2024
[25] https://github.com/DIPSA-QUB/ParaGrapher/blob/main/test/read_bandwidth.c
[26] https://github.com/DIPSA-QUB/ParaGrapher/blob/main/test/ReadBandwidth.java
[27] This has been investigated and implemented in [23].

**HDD**

| G5 | 8 ME | 64 ME | 128 ME | SH | 8 ME | 64 ME | 128 ME | CW | 8 ME | 64 ME | 128 ME | MS | 8 ME | 64 ME | 128 ME |
|---|---|---|---|---|---|---|---|---|---|---|---|---|---|---|---|
| 9 T | 488 | 493 | 509 | 9 T | 439 | 397 | 384 | 9 T | 452 | 410 | 608 | 9 T | 2976 | 2818 | 2954 |
| 18 T | 498 | 576 | 634 | 18 T | 407 | 422 | 379 | 18 T | 357 | 325 | 438 | 18 T | 2977 | 3279 | 3625 |
| 36 T | 560 | 691 | 755 | 36 T | 431 | 415 | 440 | 36 T | 360 | 374 | 369 | 36 T | 3804 | 4361 | 4553 |
| | ↓488 | ↕578 | ↑755 | | ↓379 | ↕412 | ↑440 | | ↓325 | ↕410 | ↑608 | | ↓2818 | ↕3483 | ↑4553 |

**SSD**

| G5 | 8 ME | 64 ME | 128 ME | SH | 8 ME | 64 ME | 128 ME | CW | 8 ME | 64 ME | 128 ME | MS | 8 ME | 64 ME | 128 ME |
|---|---|---|---|---|---|---|---|---|---|---|---|---|---|---|---|
| 9 T | 156 | 155 | 155 | 9 T | 344 | 340 | 499 | 9 T | 589 | 397 | 390 | 9 T | 848 | 780 | 801 |
| 18 T | 137 | 185 | 107 | 18 T | 236 | 230 | 236 | 18 T | 272 | 307 | 469 | 18 T | 469 | 444 | 467 |
| 36 T | 102 | 107 | 111 | 36 T | 238 | 226 | 233 | 36 T | 280 | 288 | 301 | 36 T | 512 | 493 | 444 |
| | ↓102 | ↕135 | ↑185 | | ↓226 | ↕287 | ↑499 | | ↓272 | ↕366 | ↑589 | | ↓444 | ↕584 | ↑848 |

**Figure 8: Execution time of ParaGrapher (in seconds) for 9, 18, and 36 threads and buffer sizes of 8, 64, and 128 Million edges**

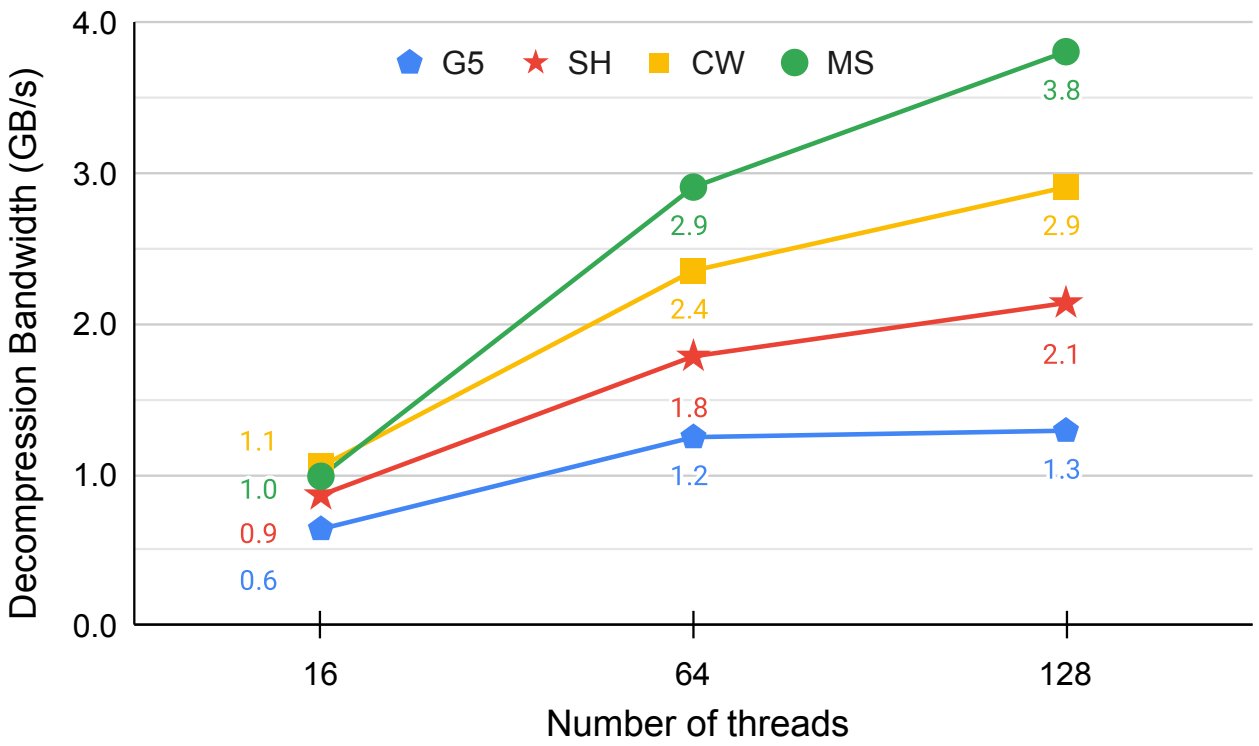

**Figure 9: Scalability of decompression bandwidth** - Datasets are stored on memory to ignore performance impacts of storage

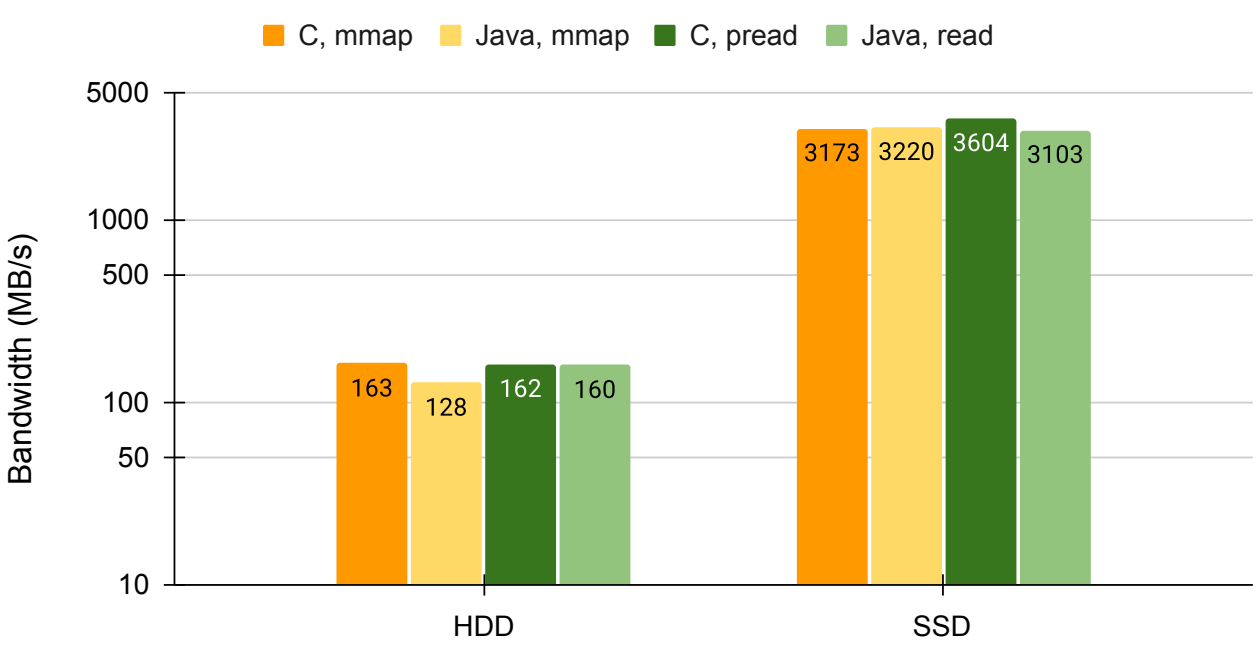

**Figure 10: Read bandwidth of Java vs C**

format is needed for graph partitioning and before graph loading, the underlying graph should be accessed to extract the offsets array. ParaGrapher stores the `offsets` array as a binary file to accelerate graph partitioning without accessing the compressed graph.

**File Size Limitation Flexibility**. As storage mechanisms (including file systems and cloud services) have theoretical and practical limits on the size of a single file/inode, it helps the end-user to store the graph dataset as multi-part files and to pass them directly to the graph loading system without merging.

**Integrity Validation**. Integrity checking can be done at different levels, including OS secure filesystems, mirroring, and software/hardware RAID. However, the graph loading process can validate the integrity of requested edge blocks. Some graph datasets provide checksum for this purpose[28] [34].

**Efficient Compression**. While efficient decompressing of large graphs is a challenge, there are cases where the efficiency of the graph compression is also a bottleneck. As an example, when the execution of graph algorithms should be temporarily suspended, efficient compression of the processing graph datasets is the key to minimising storage space [9].

**Network-Based Distributed Decompression**. In this paper, we assumed that in a distributed computation model, each machine performs the decompression independently from the other ones. Another option is to divide the decompression between machines and the results to be shared through the network. That brings network bandwidth as a new factor to the decompression efficiency model we presented in Section 3 but with an increment in the decompression bandwidth ($d$). This decompression model will be useful for distributed graph processing with overlaps on edges which are processed by different machines.

## 7 FURTHER RELATED WORK

**Graph Compression**. Graph representation may be used to identify large (bi)cliques to reduce the number of edges that should be stored [14, 49]. Graph summarization tries to create a group of vertices and then edges will be either between these group or written in a correction list [46, 61]. Lossy compression does not produce the original graph after decompression and techniques including frontier sampling [47], query-preserving compression [13], and importance-based sampling [18] may be applied. Rule-based compression can be helpful for compression techniques that work based on the similarity of neighbors [8]. We refer to compression surveys for further consideration [6, 21, 41].

**Graph Loading & Preprocessing**. PIGO[29] [17] is a parallel library for loading uncompressed graphs in COO/CSX formats. The library defines a graph structure and loads the graph completely. GVEL [52] optimizes conversion of edge lists (COO format) to CSX. In addition to graph loading, distributed-memory and streaming graph processing require further actions (such as partitioning) before performing a particular graph analytic task. EndGraph optimizes load-balance and sorting in distributed preprocessing [38]. GraPU optimizes streaming preprocessing by component classification of the impacted vertices and by analytic-based precomputing to accelerate convergence after merging [54]. LV et al. present a survey of graph preprocessing methods [39].

[28] https://blogs.qub.ac.uk/DIPSA/MS-BioGraphs-Validation/
[29] https://github.com/GT-TDAlab/PIGO

**Locality Optimizing**. Locality optimizing algorithms play an important role in improving compression ratio of those graph compression methods that try to find common patterns between neighbor lists. Different locality-optimizing algorithms have been suggested [56, 57, 60] and has been evaluated [12, 28, 29]. While dataset-based locality optimizing algorithms will also result in better performance of graph algorithms, algorithm-based locality optimisation may be used to optimize locality of memory accesses in execution of graph algorithms [22, 27, 30–33].

## 8 CONCLUSION

In this paper, we introduced ParaGrapher, an API and library to optimize loading large-scale compressed graphs. We presented a model for loading large-scale compressed graphs based on the characteristics of storage and the decompression bandwidth.

We evaluated ParaGrapher for 6 real-world and synthetic graphs with up to 124 Billion edges in comparison to uncompressed formats in a state-of-the-art graph processing framework. The evaluation shows that ParaGrapher provides up to 3.2 times speedup for loading graphs and up to 5.2 times speedup for end-to-end execution.

We hope ParaGrapher and the analysis provided in this paper will be helpful in (i) accelerating research in High-Performance Graph Processing, in (ii) facilitating a more comprehensive evaluation of contributions, and in (iii) motivating HPC research on optimizing the decompression process, especially for fast storage mediums and high-performance parallel/distributed file systems.

## SOURCE CODE AVAILABILITY

Source code of ParaGrapher is available online on https://blogs.qub.ac.uk/DIPSA/ParaGrapher/.

## ACKNOWLEDGMENTS

We are grateful to Prof. Sebastiano Vigna for his helps on this project, to Tony Lindsay for the administration of the HPDC cluster, EEECS, Queen's University Belfast (QUB), and to Northern Ireland High Performance Computing Centre (NI-HPC) team including Dr. Vaughan Purnell for supporting compute requirements.

We also appreciate the reviewers of our paper in IEEE BigData 2025 for their valuable feedback.

Working on the first versions of ParaGrapher, first author was supported by a scholarship from the Department for the Economy, Northern Ireland and QUB.

The third author is supported by a grant from the Ministry of Education, India, under a collaborative project between Tezpur University, India and QUB.

This work is partially supported by Horizon Europe under grant agreement 101072456 (RELAX), the Engineering and Physical Sciences Research Council under grant agreement EP/X01794X/1 (ASCCED), EP/X029174/1 (RELAX), EP/Z531054/1 (the Kelvin Living Lab), and EP/T022175/1 (Kelvin-2 Tier-2 HPC).

## REFERENCES

[1] Ariful Azad et al. 2020. Evaluation of Graph Analytics Frameworks Using the GAP Benchmark Suite. In *2020 IEEE International Symposium on Workload Characterization (IISWC)*. IEEE, USA, 216–227. https://doi.org/10.1109/IISWC50251.2020.00029

[2] Scott Beamer, Krste Asanovic, and David A. Patterson. 2015. The GAP Benchmark Suite. *CoRR* abs/1508.03619 (2015), 1–16. arXiv:1508.03619

[3] Ronald F Boisvert, Roldan Pozo, and Karin A Remington. 1996. *The matrix market exchange formats: Initial design*. Vol. 5935. National Bureau of Standards.

[4] Paolo Boldi, Antoine Pietri, Sebastiano Vigna, and Stefano Zacchiroli. 2020. Ultra-Large-Scale Repository Analysis via Graph Compression. In *2020 (SANER)*. IEEE Computer Society, London, ON, Canada, 184–194. https://doi.org/10.1109/SANER48275.2020.9054827

[5] Paolo Boldi and Sebastiano Vigna. 2004. The Webgraph Framework I: Compression Techniques. In *Proceedings of the 13th International Conference on World Wide Web* (New York, NY, USA) *(WWW '04)*. Association for Computing Machinery, New York, NY, USA, 595–602. https://doi.org/10.1145/988672.988752

[6] Šejla Čebirić, François Goasdoué, Haridimos Kondylakis, Dimitris Kotzinos, Ioana Manolescu, Georgia Troullinou, and Mussab Zneika. 2019. Summarizing semantic graphs: a survey. *The VLDB journal* 28 (2019), 295–327.

[7] Deepayan Chakrabarti, Yiping Zhan, and Christos Faloutsos. 2004. R-MAT: A Recursive Model for Graph Mining.. In *SDM*. SIAM, 442–446.

[8] Zheng Chen, Feng Zhang, JiaWei Guan, Jidong Zhai, Xipeng Shen, Huanchen Zhang, Wentong Shu, and Xiaoyong Du. 2023. Compressgraph: Efficient parallel graph analytics with rule-based compression. *Proceedings of the ACM on Management of Data* 1, 1 (2023), 1–31.

[9] Miaomiao Cheng, Cheng Zhao, Liang Qin, Hexiang Lin, Rong Hua Li, Guoren Wang, Shuai Zhang, and Lei Zhang. 2023. ByteGAP: A Non-continuous Distributed Graph Computing System using Persistent Memory. (2023). https://ceur-ws.org/Vol-3462/ADMS7.pdf

[10] Leonardo Dagum and Ramesh Menon. 1998. OpenMP: an industry standard API for shared-memory programming. *IEEE Computational Science and Engineering* 5, 1 (1998), 46–55. https://doi.org/10.1109/99.660313

[11] Timothy A. Davis and Yifan Hu. 2011. The university of Florida sparse matrix collection. *ACM Trans. Math. Softw.* 38, 1, Article 1 (dec 2011), 25 pages. https://doi.org/10.1145/2049662.2049663

[12] Priyank Faldu, Jeff Diamond, and Boris Grot. 2019. A Closer Look at Lightweight Graph Reordering. In *2019 IEEE International Symposium on Workload Characterization (IISWC)*. IEEE, 1–13. https://doi.org/10.1109/IISWC47752.2019.9041948

[13] Wenfei Fan, Jianzhong Li, Xin Wang, and Yinghui Wu. 2012. Query preserving graph compression. In *Proceedings of the 2012 ACM SIGMOD international conference on management of data*. 157–168.

[14] Tomás Feder and Rajeev Motwani. 1991. Clique partitions, graph compression and speeding-up algorithms. In *Proceedings of the twenty-third annual ACM symposium on Theory of computing*. 123–133.

[15] Henning Fernau and Kshitij Gajjar. 2022. The Space Complexity of Sum Labelling. arXiv:2107.12973 [cs.DS]

[16] Tommaso Fontana, Sebastiano Vigna, and Stefano Zacchiroli. 2024. WebGraph: The Next Generation (Is in Rust). In *ACM Web Conference 2024*.

[17] Kasimir Gabert and Ümit V. Çatalyürek. 2021. PIGO: A Parallel Graph Input/Output Library. In *2021 IEEE International Parallel and Distributed Processing Symposium Workshops (IPDPSW)*. IEEE, 276–279. https://doi.org/10.1109/IPDPSW52791.2021.00050

[18] Anna C Gilbert and Kirill Levchenko. 2004. Compressing network graphs. In *Proceedings of the LinkKDD workshop at the 10th ACM Conference on KDD*, Vol. 124.

[19] Siddhartha V. Jayanti and Robert E. Tarjan. 2016. A Randomized Concurrent Algorithm for Disjoint Set Union. *CoRR* abs/1612.01514 (2016). arXiv:1612.01514

[20] George Karypis and Vipin Kumar. 1995. *METIS – Unstructured Graph Partitioning and Sparse Matrix Ordering System, Version 2.0*. Technical Report.

[21] Arijit Khan, Sourav Saha Bhowmick, and Francesco Bonchi. 2017. Summarizing static and dynamic big graphs. (2017).

[22] Mohsen Koohi Esfahani. 2022. *On Designing Structure-Aware High-Performance Graph Algorithms*. Ph. D. Dissertation. Queen's University Belfast. https://blogs.qub.ac.uk/dipsa/ODSAHPGA

[23] Mohsen Koohi Esfahani. 2025. Accelerating Loading WebGraphs in ParaGrapher. *CoRR* abs/2507.2507.00716 (2025). arXiv:2507.00716

[24] Mohsen Koohi Esfahani, Paolo Boldi, Hans Vandierendonck, Peter Kilpatrick, and Sebastiano Vigna. 2023. Dataset Announcement: MS-BioGraphs, Trillion-Scale Public Real-World Sequence Similarity Graphs. In *IISWC'23* (Belgium, Ghent). IEEE Computer Society. https://doi.org/10.1109/IISWC59245.2023.00029

[25] Mohsen Koohi Esfahani, Paolo Boldi, Hans Vandierendonck, Peter Kilpatrick, and Sebastiano Vigna. 2023. MS-BioGraphs: Sequence Similarity Graph Datasets. *CoRR* abs/2308.16744 (2023). https://doi.org/10.48550/arXiv.2308.16744 arXiv:2308.16744 [cs.DC]

[26] Mohsen Koohi Esfahani, Paolo Boldi, Hans Vandierendonck, Peter Kilpatrick, and Sebastiano Vigna. 2023. On Overcoming HPC Challenges of Trillion-Scale Real-World Graph Datasets. In *2023 IEEE International Conference on Big Data (BigData'23)* (Italia, Sorrento). IEEE Computer Society. https://doi.org/10.1109/

BigData59044.2023.10386309

[27] Mohsen Koohi Esfahani, Peter Kilpatrick, and Hans Vandierendonck. 2021. Exploiting In-Hub Temporal Locality In SpMV-Based Graph Processing. In *50th International Conference on Parallel Processing* (Lemont, IL, USA) *(ICPP 2021)*. Association for Computing Machinery, New York, NY, USA, 10 pages. https://doi.org/10.1145/3472456.3472462

[28] Mohsen Koohi Esfahani, Peter Kilpatrick, and Hans Vandierendonck. 2021. How Do Graph Relabeling Algorithms Improve Memory Locality?. In *2021 IEEE International Symposium on Performance Analysis of Systems and Software (ISPASS)*. IEEE Computer Society, USA, 84–86. https://doi.org/10.1109/ISPASS51385.2021.00023

[29] Mohsen Koohi Esfahani, Peter Kilpatrick, and Hans Vandierendonck. 2021. Locality Analysis of Graph Reordering Algorithms. In *2021 IEEE International Symposium on Workload Characterization (IISWC'21)*. IEEE Computer Society, USA, 101–112. https://doi.org/10.1109/IISWC53511.2021.00020

[30] Mohsen Koohi Esfahani, Peter Kilpatrick, and Hans Vandierendonck. 2021. Thrifty Label Propagation: Fast Connected Components for Skewed-Degree Graphs. In *2021 IEEE CLUSTER*. IEEE Computer Society, USA, 226–237. https://doi.org/10.1109/Cluster48925.2021.00042

[31] Mohsen Koohi Esfahani, Peter Kilpatrick, and Hans Vandierendonck. 2022. LOTUS: Locality Optimizing Triangle Counting. In *27th ACM SIGPLAN Annual Symposium on Principles and Practice of Parallel Programming (PPoPP 2022)*. ACM, 219–233. https://doi.org/10.1145/3503221.3508402

[32] Mohsen Koohi Esfahani, Peter Kilpatrick, and Hans Vandierendonck. 2022. MASTIFF: Structure-Aware Minimum Spanning Tree/Forest. In *36th ACM International Conference on Supercomputing*. Association for Computing Machinery, New York, NY, USA, 13 pages. https://doi.org/10.1145/3524059.3532365

[33] Mohsen Koohi Esfahani, Peter Kilpatrick, and Hans Vandierendonck. 2022. SAPCo Sort: Optimizing Degree-Ordering for Power-Law Graphs. In *2022 IEEE International Symposium on Performance Analysis of Systems and Software (ISPASS)*. IEEE Computer Society. https://doi.org/10.1109/ISPASS55109.2022.00015

[34] Mohsen Koohi Esfahani, Sebastiano Vigna, Paolo Boldi, Hans Vandierendonck, and Peter Kilpatrick. 2024. MS-BioGraphs: Trillion-Scale Sequence Similarity Graph Datasets. https://doi.org/10.21227/gmd9-1534

[35] Jérôme Kunegis. 2013. KONECT: The Koblenz Network Collection. In *Proceedings of the 22nd International Conference on World Wide Web* (Rio de Janeiro, Brazil) *(WWW '13 Companion)*. Association for Computing Machinery, New York, NY, USA, 1343–1350. https://doi.org/10.1145/2487788.2488173

[36] Haewoon Kwak, Changhyun Lee, Hosung Park, and Sue Moon. 2010. What is Twitter, a Social Network or a News Media?. In *Proceedings of the 19th International Conference on World Wide Web* (Raleigh, North Carolina, USA) *(WWW '10)*. Association for Computing Machinery, New York, NY, USA, 591–600. https://doi.org/10.1145/1772690.1772751

[37] Yi Li, Tsun-Yu Yang, Ming-Chang Yang, Zhaoyan Shen, and Bingzhe Li. 2024. Celeritas: Out-of-Core Based Unsupervised Graph Neural Network via Cross-Layer Computing 2024. In *2024 IEEE International Symposium on High-Performance Computer Architecture (HPCA)*. 91–107. https://doi.org/10.1109/HPCA57654.2024.00018

[38] Tianfeng Liu and Dan Li. 2022. EndGraph: An Efficient Distributed Graph Preprocessing System. In *2022 IEEE 42nd International Conference on Distributed Computing Systems (ICDCS)*. 111–121. https://doi.org/10.1109/ICDCS54860.2022.00020

[39] Zhengyang Lv, Mingyu Yan, Xin Liu, Mengyao Dong, Xiaochun Ye, Dongrui Fan, and Ninghui Sun. 2023. A Survey of Graph Pre-processing Methods: From Algorithmic to Hardware Perspectives. arXiv:2309.07581 [cs.AR]

[40] Jasmina Malicevic, Baptiste Lepers, and Willy Zwaenepoel. 2017. Everything You Always Wanted to Know about Multicore Graph Processing but Were Afraid to Ask. In *Proceedings of the 2017 USENIX Conference on Usenix Annual Technical Conference* (Santa Clara, CA, USA) *(USENIX ATC '17)*. USENIX Association, USA, 631–643.

[41] Sebastian Maneth and Fabian Peternek. 2015. A survey on methods and systems for graph compression. *arXiv preprint arXiv:1504.00616* (2015).

[42] Sebastian Maneth and Fabian Peternek. 2018. Grammar-based graph compression. *Information Systems* 76 (2018), 19–45.

[43] Abbas Mazloumi, Mahbod Afarin, and Rajiv Gupta. 2023. Expressway: Prioritizing Edges for Distributed Evaluation of Graph Queries. In *2023 IEEE International Conference on Big Data (BigData)*. 4362–4371. https://doi.org/10.1109/BigData59044.2023.10386860

[44] Lingkai Meng, Yu Shao, Long Yuan, Longbin Lai, Peng Cheng, Xue Li, Wenyuan Yu, Wenjie Zhang, Xuemin Lin, and Jingren Zhou. 2024. A Survey of Distributed Graph Algorithms on Massive Graphs. arXiv:2404.06037 [cs.DC]

[45] Gary L. Miller, Richard Peng, and Shen Chen Xu. 2013. Parallel graph decompositions using random shifts. In *Proceedings of the Twenty-Fifth Annual ACM Symposium on Parallelism in Algorithms and Architectures* (Montréal, Québec, Canada) *(SPAA '13)*. Association for Computing Machinery, New York, NY, USA, 196–203. https://doi.org/10.1145/2486159.2486180

[46] Saket Navlakha, Rajeev Rastogi, and Nisheeth Shrivastava. 2008. Graph summarization with bounded error. In *Proceedings of the 2008 ACM SIGMOD international conference on Management of data*. 419–432.

[47] Bruno Ribeiro and Don Towsley. 2010. Estimating and sampling graphs with multidimensional random walks. In *Proceedings of the 10th ACM SIGCOMM conference on Internet measurement*. 390–403.

[48] Ryan A. Rossi and Nesreen K. Ahmed. 2015. The Network Data Repository with Interactive Graph Analytics and Visualization. In *Proceedings of the Twenty-Ninth AAAI Conference on Artificial Intelligence* (Austin, Texas) *(AAAI'15)*. AAAI Press, USA, 4292–4293.

[49] Ryan A Rossi and Rong Zhou. 2018. Graphzip: a clique-based sparse graph compression method. *Journal of Big Data* 5, 1 (2018), 10.

[50] Amitabha Roy, Ivo Mihailovic, and Willy Zwaenepoel. 2013. X-Stream: Edge-Centric Graph Processing Using Streaming Partitions. In *Proceedings of the Twenty-Fourth ACM Symposium on Operating Systems Principles* (Farminton, Pennsylvania) *(SOSP '13)*. Association for Computing Machinery, New York, NY, USA, 472–488. https://doi.org/10.1145/2517349.2522740

[51] Youcef Saad. 1994. Sparskit: a basic tool kit for sparse matrix computations - Version 2.

[52] Subhajit Sahu. 2023. GVEL: Fast Graph Loading in Edgelist and Compressed Sparse Row (CSR) formats. arXiv:2311.14650 [cs.PF]

[53] Peter Sanders and Matthias Schimek. 2023. Engineering Massively Parallel MST Algorithms. arXiv:2302.12199 [cs.DC]

[54] Feng Sheng, Qiang Cao, Haoran Cai, Jie Yao, and Changsheng Xie. 2018. GraPU: Accelerate Streaming Graph Analysis through Preprocessing Buffered Updates. In *Proceedings of the ACM Symposium on Cloud Computing* (Carlsbad, CA, USA) *(SoCC '18)*. Association for Computing Machinery, New York, NY, USA, 301–312. https://doi.org/10.1145/3267809.3267811

[55] Julian Shun, Guy E. Blelloch, Jeremy T. Fineman, Phillip B. Gibbons, Aapo Kyrola, Harsha Vardhan Simhadri, and Kanat Tangwongsan. 2012. Brief announcement: the problem based benchmark suite. In *Proceedings of the Twenty-Fourth Annual ACM Symposium on Parallelism in Algorithms and Architectures* (Pittsburgh, Pennsylvania, USA) *(SPAA '12)*. Association for Computing Machinery, New York, NY, USA, 68–70. https://doi.org/10.1145/2312005.2312018

[56] Sanya Srivastava and Tyler Sorensen. 2023. Degree-Aware Kernel Mapping for Graph Processing on GPUs. In *2023 IEEE International Symposium on Performance Analysis of Systems and Software (ISPASS)*. 319–321. https://doi.org/10.1109/ISPASS57527.2023.00039

[57] Jiawen Sun, Hans Vandierendonck, and Dimitrios S. Nikolopoulos. 2018. VEBO: A Vertex- and Edge-Balanced Ordering Heuristic to Load Balance Parallel Graph Processing. *CoRR* abs/1806.06576 (2018), 1–13. arXiv:1806.06576

[58] Michael Sutton, Tal Ben-Nun, and Amnon Barak. 2018. Optimizing parallel graph connectivity computation via subgraph sampling. In *2018 IEEE International Parallel and Distributed Processing Symposium (IPDPS)*. IEEE, 12–21. https://doi.org/10.1109/IPDPS.2018.00012

[59] Xianghao Xu, Fang Wang, Hong Jiang, Yongli Cheng, Dan Feng, and Peng Fang. 2024. A disk I/O optimized system for concurrent graph processing jobs. *Frontiers of Computer Science* 18, 3 (2024), 183105.

[60] Tianyi Zhang, Aditya Desai, Gaurav Gupta, and Anshumali Shrivastava. 2023. HashOrder: Accelerating Graph Processing Through Hashing-based Reordering. (2023).

[61] Fang Zhou. 2015. Graph compression. *Department of Computer Science and Helsinki Institute for Information Technology HIIT* (2015), 1–12.

# A APPENDIX - API DOCUMENTATION

In this section, we present a high-level review of the functions in ParaGrapher API. A detailed API documentation is provided in our GitHub repository, https://github.com/DIPSA-QUB/ParaGrapher/wiki/API-Documentation.

Most functions have two arguments `void**` **args** and `int` **argc** that have been used to pass additional arguments (for sending input to the library or receiving data from library) that may be required for particular graph types.

## A.1 Initialization

```
int paragrapher_init();
```

This function is essential to start using the library. Upon calling this function, ParaGrapher iterates over its inner files that have implemented the API for loading different graph formats and creates a list of their functions which is used for accessing graphs. The return value is either `0` on success, or `-1` on failure.

## A.2 Opening A Graph

```
paragrapher_graph* paragrapher_open_graph(
  char* name, paragrapher_graph_type type,
  void** args, int argc);
```

This function opens a graph specified by filename and type (Table 2).

## A.3 Getting Info and Setting Configuration

```
int paragrapher_get_set_options(
  paragrapher_graph* graph,
  paragrapher_request_type request_type,
  void** args, int argc);
```

This function configures the settings or retrieves configurations such as library parameters (e.g., buffer size and number of buffers that control parallelism), number of vertices and edges in the graph, and status of a read request.

## A.4 Accessing Offsets and Weights of Vertices

```
void* paragrapher_csx_get_offsets(
  paragrapher_graph* graph,
  void* values,
  unsigned long start_vertex,
  unsigned long end_vertex,
  void** args, int argc);
```

```
void* paragrapher_csx_get_vertex_weights(
  ...);
```

The above functions access vertex offsets and weights. The signature of both functions is the same.

```
void
paragrapher_csx_release_offsets_weights_arrays(
  paragrapher_graph* graph, void* array);
```

By calling this function, user informs ParaGrapher that `array` (containing vertex weights/offsets) is no longer needed.

## A.5 Loading Edges of a CSX Graph

```
typedef void (*paragrapher_csx_callback)(
  paragrapher_read_request* request,
  paragrapher_edge_block* eb,
  void* offsets, void* edges,
  void* buffer_id, void* args);
```

The `callback` function is defined by the user and called by the ParaGrapher in loading a subgraph asynchronously after decompressing each block of edges. This way, the user will start processing the loaded data without waiting for completion of loading the whole requested subgraph/graph.

```
paragrapher_read_request* csx_get_subgraph(
  paragrapher_graph* graph,
  paragrapher_edge_block* eb,
  void* offsets, void* edges,
  paragrapher_csx_callback callback,
  void* callback_args, void** args, int argc);
```

This function starts loading a CSX graph or its subgraph (specified by `eb`). Depending on the implementation, load can be done synchronously (i.e., the `edges` array is filled by the library) or asynchronously (i.e., the `callback` function is called multiple times to pass read block of edges to user).

```
void csx_release_read_buffer(
  paragrapher_read_request* graph,
  paragrapher_edge_block* eb,
  void* offsets, void* edges);
```

This function is used at the end of `callback` function to inform the library that the buffer will not be used any further and its memory can be reused by the library.

```
void csx_release_read_request(
  paragrapher_read_request* request);
```

This function releases all resources associated with a returned `paragrapher_read_request` and should be called upon completion of the load process.

## A.6 Loading Edges of a COO Graph

```
paragrapher_read_request* coo_get_edges(
  paragrapher_graph* graph,
  unsigned long start_row,
  unsigned long end_row,
  void* edges,
  paragrapher_coo_callback callback,
  void* callback_args, void** args, int argc);
```

This function is similar to `csx_get_subgraph()` but for loading a COO graph or its subgraph (specified by `start_row` and `end_row`) synchronously or asynchronously.

### A.7 Releasing Graph

```
int paragrapher_release_graph(
  paragrapher_graph* graph,
  void** args, int argc)
```

This function releases the resources allocated by the library for accessing a graph and should be called as the last step of accessing/loading a graph.